# Towards Distributed Clouds

A review about the evolution of centralized cloud computing, distributed ledger technologies, and a foresight on unifying opportunities and security implications

Magnus Westerlund
Arcada University of Applied Sciences, Finland
Email: magnus.westerlund@aracada.fi

Nane Kratzke
Lübeck University of Applied Sciences, Germany
Email: nane.kratzke@fh-luebeck.de

*Abstract*—This review focuses on the evolution of cloud computing and distributed ledger technologies (blockchains) over the last decade. Cloud computing relies mainly on a conceptually centralized service provisioning model, while blockchain technologies originate from a peer-to-peer and a completely distributed approach. Still, noteworthy commonalities between both approaches are often overlooked by researchers. Therefore, to the best of the authors knowledge, this paper reviews both domains in parallel for the first time. We conclude that both approaches have advantages and disadvantages. The advantages of centralized service provisioning approaches are often the disadvantages of distributed ledger approaches and vice versa. It is obviously an interesting question whether both approaches could be combined in a way that the advantages can be added while the disadvantages could be avoided. We derive a software stack that could build the foundation unifying the best of these two worlds and that would avoid existing shortcomings like vendor lock-in, some security problems, and inherent platform dependencies.

## I. INTRODUCTION

Over the last decade Cloud Service Providers (CSP) improved their infrastructures to host and operate cloud-native applications (CNA) in a pragmatic manner and a web-scale manner. This development has led to horizontally scalable system architectures such as microservices, corresponding deployment units like containers or unikernels, and recently serverless architecture approaches. The commonality in these designs is to make use of conceptually centralized computing, provided by public or private CSPs. This makes these kinds of architectures inherently prone to vendor lock-in [1]. Currently, a consideration for companies using CSP services is the implementation of multi-cloud hosted solutions, a design architecture were more than one CSP host a service. In this paper we consider a path forward and elaborate on a definition for distributed clouds and their service provision offerings in relation to incumbent solutions. The distributed cloud refers to a software architecture based on Distributed Ledger Technology (DLT), such as blockchains, to achieve an agnostic hosting medium for DApps. Our contribution is in the identification of a potential paradigm shift in the architectural design of the distributed cloud and the mechanisms enabling a shift towards distributed clouds.

The **outline** of this paper is structured as follows. We provide a short review of cloud-native designs over the last 10 years in **Section II**, to get a better understanding of the trends in current cloud engineering. One noteworthy consideration is the vendor lock-in dilemma [1], [2]. Once a cloud application is deployed to a cloud infrastructure it is often inherently bound to that (conceptually centralized) infrastructure [3]. This is the main downside of conceptually centralized service concepts and here distributed ledger technologies might provide new solutions. Therefore, **Section III** will investigate how distributed ledger technologies are maturing and categorizes three generational shifts that might provide solutions to overcome the lock-in aspect. **Section IV** illustrates how all this could be integrated into a consistent distributed cloud concept and how distributed ledger technologies could be used as software connectors for CSP independent and distributed services [4]. We will summarize our considerations in **Section V** to derive some conclusions regarding unifying opportunities and security implications.

## II. REVIEW OF CLOUD NATIVE DESIGNS

Cloud computing emerged some 10 years ago to much chagrin of privacy conscious developers. In the first adoption phase it was likely that existing IT systems were simply transferred to cloud environments without changing the original design and architecture of these applications. Tiered applications have simply been migrated from dedicated hardware to virtualized hardware in the cloud. However, cloud environments are elastic. Elasticity is understood as the degree to which a system adapts to workload changes by provisioning and de-provisioning resources automatically. Over time, system engineers learned to understand the elasticity options of modern cloud environments better. Eventually, systems were designed for such elastic cloud infrastructures, which increased the utilization rates of underlying computing infrastructures via new design approaches like microservices or serverless architectures. This design intention is often expressed using the term cloud-native. Figure 1 shows an observable trend over the last decade.

It is often said that cloud-native applications are intentionally designed for the elastic cloud. Although this understanding can be broadly used, it does not guide and explain what a cloud-native application is exactly.

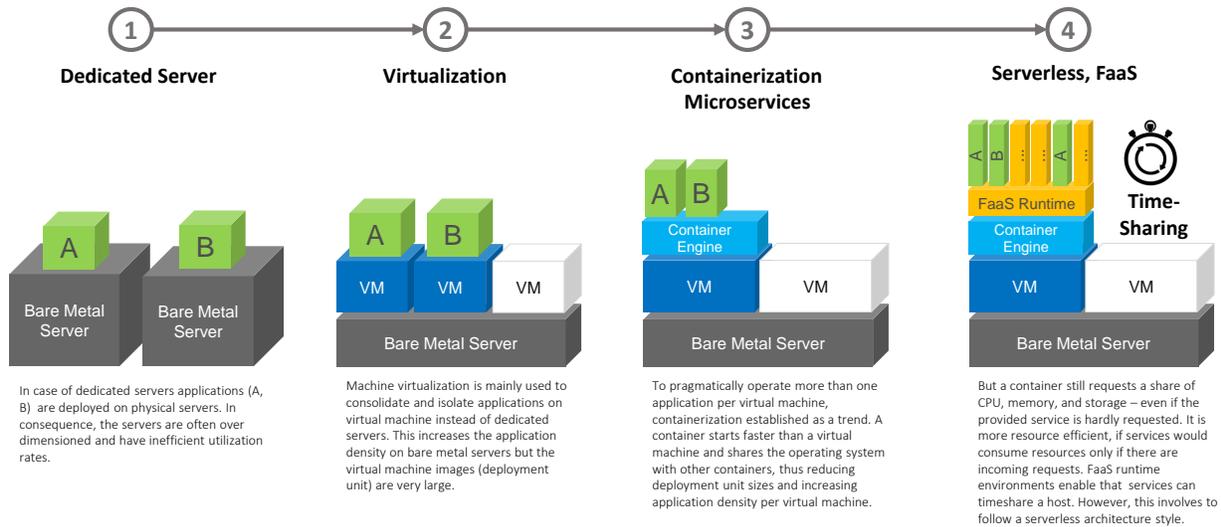

Fig. 1. The serverless "layer explosion" (Machine, OS virtualization, and FaaS runtime technology layers are stacked to optimize resource utilization)

TABLE I
OBSERVABLE CNA SOFTWARE ENGINEERING TRENDS

| Trend | Rationale |
| --- | --- |
| Microservices | Microservices can be seen as a "pragmatic" interpretation of SOA. In addition to SOA microservice architectures intentionally focus and compose small and independently replaceable horizontally scalable services that are "doing one thing well". [5], [6] |
| DevOps | DevOps is a practice that emphasizes the collaboration of software developers and IT operators. It aims to build, test, and release software more rapidly, frequently, and more reliably using automated processes for software delivery. |
| Softwareization | Softwareization of infrastructure and network enables to automate the process of software delivery and infrastructure changes more rapidly. |
| Standardized Deployment Units | Deployment units wrap a piece of software in a complete file system that contains everything needed to run: code, runtime, system tools, system libraries. This guarantees that the software will always run the same, regardless of its environment. This is often done using container technologies (OCI standard). Unikernels would work as well but are not yet in widespread use. A deployment unit should be designed and interconnected according to a **collection of cloud-focused patterns** like the *twelve-factor app* collection [7], the *circuit breaker* pattern [8] or *cloud computing patterns* [9], [10]. |
| Elastic Platforms | Elastic platforms like Kubernetes, Mesos, or Swarm can be seen as a unifying middleware of elastic infrastructures. Elastic platforms extend resource sharing and increase the utilization of underlying compute, network and storage resources for custom but standardized deployment units. |
| Serverless | The term serverless is used for an architectural style that is used for cloud application architectures that deeply depend on external third-party-services (Backend-as-a-Service, BaaS) and integrating them via small event-based triggered functions (Function-as-a, FaaS). FaaS extend resource sharing of elastic platforms by simply by applying time-sharing concepts. |
| State Isolation | Stateless components are easier to scale up/down horizontally than stateful components. Of course, stateful components can not be avoided, but stateful components should be reduced to a minimum and realized by intentional horizontal scalable storage systems (often eventual consistent NoSQL databases). |
| Versioned REST APIs | REST-based APIs provide scalable and pragmatic communication, means relying mainly on already existing internet infrastructure and well defined and widespread standards. |
| Loose coupling | Service composition is done by events or by data. Event coupling relies on messaging solutions (e.g. AMQP standard). Data coupling relies often on scalable but (mostly) eventual consistent storage solutions (which are often subsumed as NoSQL databases). |

CNA, thus, refers to an unconsolidated software engineering domain with isolated engineering trends (see Table I). Although there is no common definition that explains what a CNA exactly is, there is a common but unconscious understanding across several relevant studies. Fehling et al. propose that a cloud-native application should be IDEAL, so it should have an **[i]solated state**, is **[d]istributed** in its nature, is **[e]lastic** in a horizontal scaling way, is operated via an **[a]utomated management** system and its components should be **[l]oosely coupled** [9]. According to Stine [11] there are common motivations for cloud-native application architectures, e.g. to deliver software-based solutions quicker **(speed)**, in a more fault isolating, fault tolerating, and automatic recovering way **(safety)**, to enable horizontal (instead of vertical) application scaling **(scale)**, and finally to handle a huge diversity of (mobile) platforms and legacy systems **(client diversity)**.

Microservice architectures make use of containers as deployment units. Because containers are much more lightweight than virtual machines, many more containers can be deployed on a physical machine. Hence, microservices increase application density on computing infrastructures.

A new architecture "serverless" is getting more and more common since 2014. Serverless is implemented using Function-as-a-Service (FaaS) concepts [12]. Figure 1 shows the evolution of how resource utilization has been optimized over the last 10 years ending in the latest trend to make use of Faas platforms in serverless architectures. FaaS platforms

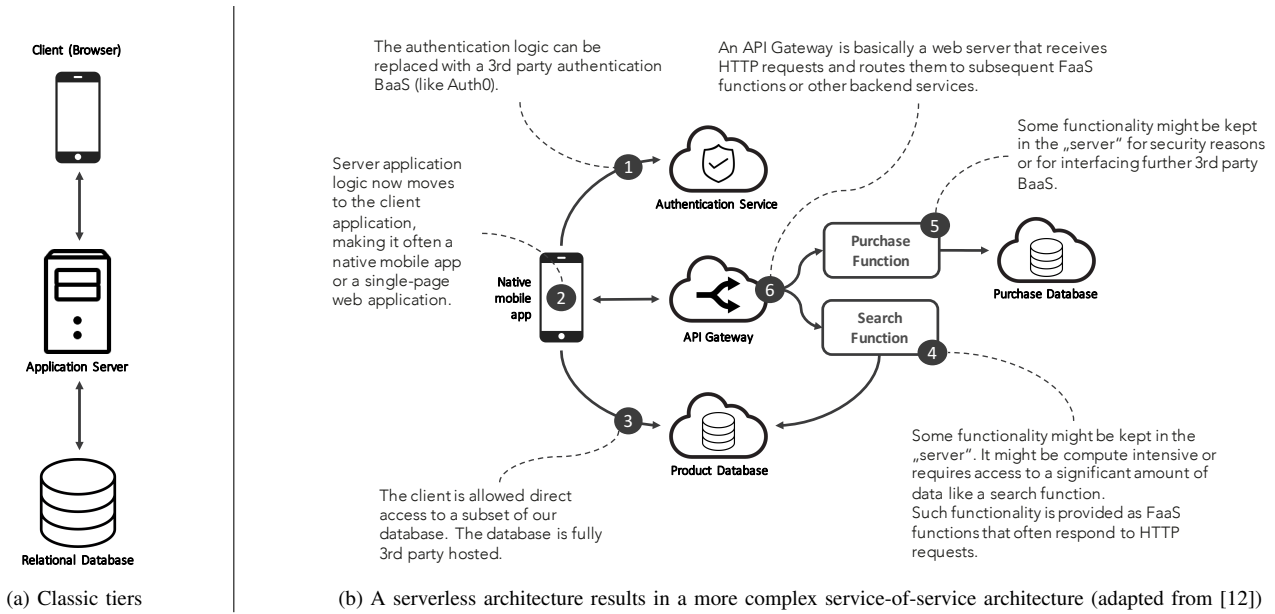

Fig. 2. From n-tiered to serverless architectures (complexity that has been encapsulated and hidden in a monolithic application server now is brought to light)

apply time-sharing principles and increase the utilization factor of computing infrastructures, and thus avoid expensive always-on components. Studies showed, that due to this time-sharing, serverless architectures can reduce costs by 70% [13]. However, Figure 2 shows that serverless architectures (and microservice architectures as well) require a cloud application architecture redesign, compared to classical e-commerce applications.

FaaS computing provides some inherent benefits like resource and cost efficiency, operation simplicity, and a possible increase of development speed and improved time-to-market [12]. According to [12] FaaS computing comes along with some noteworthy drawbacks, like runtime constraints (**R1**), state constraints (**R2**) and unsolved function composition problems (**R3**). Resulting serverless architectures have security implications (**O1**). They increase likely attack surfaces and shift part of the application logic to the client-side (which is not under complete control of service providers). Furthermore, FaaS increases vendor lock-in problems (**O2**), client complexity (**D1**), and integration and testing complexity (**D2**). Table II summarizes the benefits and drawbacks of FaaS.

### III. EVOLUTION OF DISTRIBUTED LEDGER TECHNOLOGY

In the previous section we reviewed the current centralised CNA offered by CSPs. In this section we look towards the future and consider what DLT will mean for the cloud computing landscape. DLT has become a term for describing different implementations of a distributed ledger-based consensus mechanism, including the original blockchain. There are two main categories of achieving consensus that relates to who can participate in the consensus process. A DLT is either considered permissioned based (private) or permissionless (public), this paper focuses on public DLTs. Still, many of the findings are relevant for private DLTs as well.

We first review the 10 year history of DLT (incl. blockchain technology) and current proposals for DLT development. By grouping findings into three generational shifts we determine what features have enabled new innovative views. The generational grouping is based on a mix of primary (academic) and secondary (e.g. white papers) sources. The grouping for the third generation is mostly based on secondary sources such as white papers and presentations by development groups.

#### A. First generation blockchain

The Bitcoin blockchain introduced a cryptographically secured and distributed ledger [15]. The ability to append transactions to an otherwise immutable ledger is based on a distributed and pseudonymous consensus mechanism, i.e. Nakamoto consensus. Bitcoin's consensus protocol includes both a validity check of a certain transaction and an information sharing protocol, where accepted transactions are stored in blocks that in turn are chained together in a chronological order. A distinction of blockchains compared to e.g. a relational database, is that the ledger is a transactional database. Thus, the blockchain only stores transactional changes and thereby stays immutable by not forcing an update on pre-existing variable values. Here we should mention that transactional databases that are not based on DLT, also exist. For a Bitcoin blockchain overview see [16] and [17].

Several other cryptocurrency blockchains were introduced that mirrored Bitcoin's Proof of Work (PoW) mining protocol, that is used for mining new blocks on the chain and for deterring attacks on the chain [18]. The cryptographic complexity in calculating the following block, in regards to Bitcoin's PoW algorithm, is currently configured so that a new block is created roughly every 10 minutes [19]. This means that for a transaction to be validated and thereby included in a block, the latency is roughly 10min. However due to the double-spend vulnerability it is considered prudent to wait for another 5 blocks (60min latency) to be created in order to certify that the original transaction was not double-spent [19]. There are other factors to consider for defining transaction latency as well, but here we refer to an optimal situation. The computing complexity of Bitcoin's PoW is reset every 2016 blocks (about two weeks), to adapt to variation in mining power, while maintaining the block renewal time [20]. For other blockchains the block renewal process is often set to a shorter timespan. Litecoin aims to produce a new block every 2.5 minutes and Ethereum around 15s [18]. In combination with block size the renewal time influences the maximum number of transactions per second.

TABLE II
FaaS BENEFITS AND DRAWBACKS REPORTED BY PRACTITIONERS (LIST LIKELY NOT COMPLETE, MAINLY COMPILED FROM [12])

| Benefits | Drawbacks |
|---|---|
| **RESOURCE EFFICIENCY (COST)**<br>- auto-scaling based on trigger stimulus<br>- reduced operational costs<br>- time-sharing (no always-on components necessary) | **R1: RUNTIME CONSTRAINTS AND EFFECTS**<br>- maximum function runtime is limited (due to API gateway timeouts, function timeouts)<br>- startup latencies must be considered (especially for complex runtime environments like Java)<br>- function runtime variations (plenty of influencing performance factors out of direct control)<br>**R2: STATE CONSTRAINTS**<br>- functions can not preserve a state across function calls<br>- external state (cache, key/value stores, etc.) can compensate this but is a magnitude slower<br>**R3: FUNCTION COMPOSITION PROBLEMS**<br>- unresolved function composition problems, like the serverless trilemma (see [14])<br>- and double spending problems if FaaS functions call other FaaS functions synchronously. |
| **OPERATION** (service side)<br>- simplified deployment<br>- simplified operation (see auto-scaling) | **O1: SECURITY**<br>- increased attack surfaces (service level, each service endpoint introduces possible vulnerabilities)<br>- increased technology layers (technology stack, each layer introduces possible vulnerabilities)<br>- missing protective barrier of a monolithic server application<br>- parts of the application logic are shifted to the client-side (that is not under control of the service provider)<br>**O2: VENDOR CONTROL**<br>- no FaaS standards for API gateways and FaaS runtime environments<br>- conceptually centralized<br>- increased vendor lock-in |
| **DEVELOPMENT SPEED** (service side)<br>- development speed<br>- simplified unit testing of stateless FaaS functions<br>- better time to market | **D1: CLIENT COMPLEXITY**<br>- increased client complexity (application logic is shifted to the client-side)<br>- code replication on client side across client platforms<br>- control of application workflow on client side to avoid double-sending problems of FaaS computing<br>**D2: INTEGRATION TESTING COMPLEXITY**<br>- increased integration testing complexity<br>- missing integration test tool-suites |

Based on this discussion we want to highlight the key characteristics of first generation (1G) blockchain technology:

**Positives:** PoW guarantees against attacks, the more difficult the proof becomes, the better it is considered to guard against attacks. Therefore, 1G blockchains are relatively secure, although most proofs used in 1G may be sensitive to future quantum computing attacks. The blockchain has a rather efficient data structure for storing meta-data about transactions and wallets. However as transactions start to make use of payload data, this combined with usage growth, may lead to the ledger growing rapidly in size. One of the main benefits is the concept of an immutable data storage. 1G blockchains prove the concept of cryptocurrencies and their practical use. Due to the inherent economic incentives for miners to mine, nodes are relatively well distributed, at least for the Bitcoin blockchain.

**Negatives:** Individual private keys used to sign transactions may be hacked, e.g. a private key created from a flawed random generator. Transaction delays can lead to high latency for counter party transaction validation. In case of fraud or other crimes, data cannot be removed or changed even if ordered by a court of law. Most 1G protocols cannot properly handle micro transactions, either because of latency issues or high transaction costs, or both. Protocol changes often require a hard-fork, with the risk of chain splits (some miners decide to keep the old chain running by refusing to update). Incentive and governance issues that e.g. include a centralization of mining resources when most miners join mining pools that centralize the control structure. Difficult to determine the incentive for developing the protocol, except an indeterminable future value proposition.

**Summary on 1G blockchains:**
- Stores transactions on the chain, permits additional payload data. Uses a PoW algorithm.
- Unsuitable for high-throughput solutions, due to high latency and/or high transaction costs.
- Considered relatively safe, but not quantum proof without hard-fork update.

*B. Second generation blockchain*

The introduction of the Ethereum Virtual Machine (EVM), as a runtime environment for smart contracts, created what can be considered the second generation blockchain [21]. Although the smart contract concept was already introduced by [22], as "a computerized transaction protocol that executes the terms of a contract", the EVM was the first to realise it as a layer on top of a distributed ledger.

The EVM is a Turing complete environment and can be programmed with a de facto standardized language called Solidity [23]. Currently the EVM environment is still rather limited compared to e.g. the Java programming language and the Java VM, but basic functionality for implementing logic is included. A noticeable deficiency of the environment is the ability to perform verification. The expressive nature of a smart contract transaction on Ethereum is implemented using an imperative language, which means that the result of the transaction can only be known after publication on the ledger. This is the opposite of Bitcoins declarative transaction where the resulting state is known before publication [24]. A recent study performing static symbolic analysis (e.g. considering the sequential flow in the contract) of nearly 1M smart contracts, flagged 34,200 (2,365 distinct) contracts as potentially buggy [25]. These contracts carried directly the equivalent of millions of dollars worth of Ether, but as the researchers highlight, when the contracts interact with other contracts (can either be called or call themselves), the funds at risk may potentially be much higher. Currently a new and potentially more secure language called Vyper[1] is being developed by the Ethereum community. Although its future uptake remains unknown, Vyper identifies a potential solution for more secure practices in smart contract design.

As earlier stated for 1G blockchains, a blockchain is essentially a distributed transaction database that stores new events related to one or more account addresses. Ethereum extends the 1G blockchain and introduces two types of addresses: accounts and contracts. Both are the same size, a 40 character hex string. The difference is that contract addresses have a piece of compiled code associated with it. Therefore, when a contract is initiated by a corresponding transaction to the contract address the compiled contract is executed on the EVM.[2]

A smart contract implementation is often referred to as either stateless or stateful. The stateless contract is a design pattern aimed at reducing the cost (gas) for its execution. As the contract is executed by miners a fee is paid for the transaction and at times of congestion this fee may make the execution of a contract valuable. A stateless contract does not store a state (i.e. variables and loops) nor does it fire any events. Instead stateless contracts are used in combination with a backend.

Stateful smart contracts can further be divided into two main categories self-contained and oracle dependent. Self-contained contracts refer to a design where required data is passed along with a method call and storage is handled internally on the EVM or the blockchain. The recommended design of a stateful contract is usually to make it self-contained. Oracle dependent contracts have an external dependency to a data source or processing entity, and can therefore enable complex system designs [26]. As the blockchain cannot access external data directly, an oracle is designed to provide the needed data for processing.

Although smart contracts are autonomous once deployed on the blockchain, their true utility is achieved when connected to a front-end that allows users to interact with them. For this purpose decentralized applications (DApps) exist. A DApp is essentially a front-end (HTML + JS), a blockchain RPC-connector, and the smart contracts. Different solutions exist for connecting to the blockchain, the usually recommended path for DApps on Ethereum is to install the MetaMask browser extension. The extension can inject the Ethereum Web3 API into the frontend JavaScript context, so that the DApp can communicate with the blockchain. As the MetaMask extension also includes a wallet, payments can be handled through the extension as well.

A similar workflow can also be implemented for other applications, such as for mobile apps. The interaction with deployed contracts using the address of the contract itself and a backend through which to access Ethereum, can be implemented through a RPC-backend that attaches to an existing Ethereum node over a standard protocol such as IPC, HTTP or WebSockets[3].

**Positives:** 2G blockchains provide a Turing complete contract environment (although lacking many basic functions) that enables trustless validation of smart contracts by unknown peers. Oracle implementations allow for data exchange with the blockchain through a validated API-like interface. Smart contract design patterns for versioning and discarding contracts enable a more DevOps-like style of development. The ownership and roles for contracts can be defined. Contracts are compiled and can therefore be obfuscated, thus offering some intellectual property (IP) protection. Transfers between coin and token is relatively easy to implement, inter-chain transfers are currently not directly possible.

**Negatives:** PoW is too expensive for recording smart contract transactions in terms of computation and latency. It is difficult to implement certain basic constructs, e.g. a random generator. Currently, the best option may be to use oracle based services. Bad design choices can be difficult to rectify if a contract killing mechanism is not implemented intentionally. More complex contracts are executed inefficiently and there is no inherent validation mechanism of transfer completeness. Data storage on the blockchain from smart contracts will rapidly grow the ledger size. A hard-fork of either the VM or the ledger may render existing smart contracts inoperable. The ease of creating new tokens through smart contracts and the creation of massive ICOs (initi al coin offerings) that hold large amounts of coins, means the volatility of the coin value is impacted when ICOs are created and when ICOs revert back to fiat. This may impact the sustainability of the platform ecosystem.

**Summary on 2G blockchains:**
- Stores smart contracts on the chain and evaluates them autonomously.
- Limited utility of smart contracts due to the use of PoW consensus, high-throughput or massively scalable contract applications are not viable.
- Atomic cross-chain transfers of value is currently not possible, but transfers of value between coin and token on the same ledger is possible.

---

[1]See Vyper documentation, https://github.com/ethereum/vyper

[2]For further details see the Ethereum Contract ABI, https://github.com/ethereum/wiki/wiki/Ethereum-Contract-ABI/

[3]For further details see the Ethereum JavaScript RPC-connector API, https://github.com/ethereum/web3.js

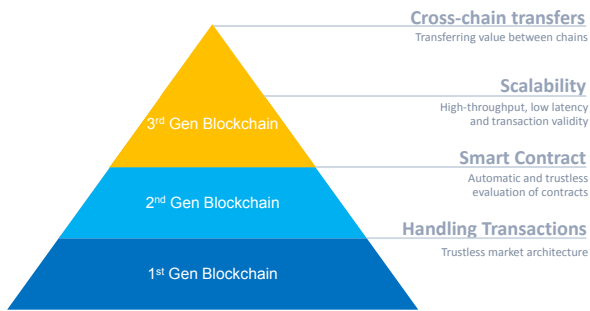

Fig. 3. Blockchain generations

- Significant issues with securing and validating intended correctness of smart contract designs.

*C. Third generation blockchain*

The 3G blockchain, or rather 3G distributed ledger technology as allowing for more diverse ledger implementations, addresses many of the limitations previously discussed. However, they also introduce fundamental problems related to centralized consensus participation [27]. We should note that 3G DLT also includes purely cryptocurrency DLTs (3G cryptocurrencies) that has not included a VM for running smart contracts. Their significance for 3G DLT stems out of the ability to perform micro transactions through e.g. Decentralized APIs (DAPI) that can be integrated into a website, DApp, or native app[4]. Currently DAPI technology is in the early development stage and is neither in widespread use nor standardized. The core building blocks for 3G DLT are to improve scalability and enable cross-chain transfers, see figure 3 for an illustration of the generations.

Scalability can be divided into the following transaction problems: throughput, latency, and validity (e.g. solving the double spend issue) [24] [28]. Throughput usually refers to the number of transactions the DLT can handle per second. For block-based chains this involves how many transactions can fit into a certain block. Latency is calculated from the initialization of a transaction until the receiver can validate its formal immutable existence. As discussed for 1G, for block-based chains this involves the renewal cycle of new blocks. For solving both the throughput and latency problems new consensus seeking mechanisms are being considered.

Increasing throughput is currently being investigated from two perspectives consensus protocol and sharding. As an example, at the time of writing a new version of Ethereum is expected to introduce a new protocol that migrates from the PoW algorithm to first a hybrid algorithm of PoW and Proof of Stake (PoS), and later to an only PoS algorithm[5]. PoS has different implementations, but a core characteristic is that those who verify transactions do so by staking their own coin value [27]. To limit the introduction of malicious nodes some kind of reputation system or stake size requirement is often used to incentivise stake nodes to be honest. A so called distributed PoS requires only a part of the stake nodes to participate at each step in the process and this group can be chosen randomly.

The Dash cryptocurrency has a version of a 2-level consensus mechanism with master nodes and miners[6]. Here master nodes take care of validating transactions and reserving the change to the wallet address, while miners then perform work to store this information on-chain. This has the benefit of increasing throughput, reducing latency (to a function of propagating information among master nodes), and eliminating the double-spend issue as the transfers are reserved as they get validated.

This shift in consensus mechanism to PoS would mean marked changes to the ecosystem as miners are not needed to calculate the next block hash. PoS for Ethereum could mean that nodes with enough coin (est. 1000 ETH initially) would be verifying the validity of transactions. As a result this may also mean that coin minting will stop for DLTs when they migrate to a more energy efficient consensus mechanism, the transaction fees are likely to be reduced as well.

Sharding the blockchain has been proposed as the long-term solution for increasing throughput [29]. The first step will be to split the blockchain network horizontally and process transactions in each 'zone'. If a transaction needs data from another zone a relay can be used to access the data. Secondly, vertical scaling of blockchains allows child chains to be spawned on top of the root. The children will perform most of the computations and communicate to the root on a regular interval. The less computations and transactions that need to be handled by the root, i.e. the more that can be scaled out to child blockchains, the higher the throughput. Although the concept has shown promise for micro transactions, it still is unknown how well security can be guaranteed on child networks and how well EVM state transitions will scale.

The third scalability problem listed above, transaction validity, will be a more difficult problem to solve. Transaction validity can be seen from a number of viewpoints, such as the programmers (code correctness), end-user (irrevocable transactions and counter party trust), seller (double spend and customer privacy), law enforcement (forensic investigation), or regulator (governing laws) to mention a few. The challenges with DLT are often determined from the perspective of how the corresponding centralized version functions and the centralized control it offers. An ideal for DLT developers is that a transaction should always be final, or "Code is law" as proposed by [30]. Without debating the correctness of the ideal, there are still numerous improvements that should be researched and that can add more security and better practices to the field [24].

Cross-chain transfers is a different way of scaling and imply that a transaction can autonomously be exchanged between two or more chains. This is particularly important for transferring value between coins, i.e. cryptocurrencies. Two different proposals has been put forward as potential solutions. First a direct transfer using a multi-signature algorithm that produces a joint signature between involved parties before being recorded on the chain. The second proposal is a trusted intermediate chain providing an escrow service, whereby each participant pay the amount into escrow that in turn releases the coins when the transaction is verified[7].

---

[4] For information see Dash Evolution 1.0 release, https://github.com/dashpay/dash-roadmap/blob/master/README.md

[5] Ethereum - Proof of Stake, https://github.com/ethereum/wiki/wiki/Proof-of-Stake-FAQ

[6] See Dash documentation, https://docs.dash.org/en/latest/masternodes/index.html

**Positives:** The development of third generation DLTs are focused on improving utility. The shift away from PoW mining to PoS validation, or other similar consensus protocols, means DLT becomes significantly more energy efficient. Proposals for scalable DLTs are promising a significant rise in throughput. The Dash cryptocurrency (among others) has shown that the double spend issue can be solved and that latency can be reduced to mere seconds through a 2-layer consensus network. Cross-chain transfers are getting new promising solutions, e.g. Cosmos, that may become an inter-chain protocol for automatic transaction conversion between chains.

**Negatives:** Many of the proposed solutions to scalability issues will incur a more centralized consensus participation, remains to be seen if the incentives are enough to keep nodes honest. There are still open questions about loyalty incentives for keeping stake nodes honest. Scalability has yet to reach a level that is sufficient for global widespread use of DLT as the underlying fabric for distributed cloud solutions. Proposals for sharding may solve throughput scalability eventually, but sharding will increase network complexity.

**Summary on 3G blockchains:**
- Third generation distributed ledger technology enables its widespread global use.
- Development work for increasing scalability, here including the sub-problems throughput, latency, and validity, is a main focus for many DLT projects and several have presented implementations.
- It remains to be seen which method cross-chain transfers will revert to, user-friendliness and sufficient security are likely key characteristics.

## IV. A SOLUTION STACK FOR THE DISTRIBUTED CLOUD

In Section II we reviewed the conceptual centralized cloud-native application evolution and in Section III we examined the evolution of distributed ledger technologies over the last decade. Both approaches have advantages and disadvantages. For instance, the centralized FaaS approach has advantages regarding performance and (fine-grained) scalability but comes with downsides regarding security, state, and overall application complexity. DLT has inherent advantages regarding security but comes with downsides regarding transaction performance and scalability. So, the advantages of centralized service provisioning approaches are often the disadvantages of DLT approaches and vice versa. It is obviously an interesting question whether both approaches could be combined in a way that the advantages can be added while the disadvantages could be avoided. This paper will not answer this question, but to start this investigation, the following Section IV-A derives a layer stack for distributed applications that takes the identified advantages and disadvantages into account.

[7]https://en.bitcoin.it/wiki/Atomic_cross-chain_trading

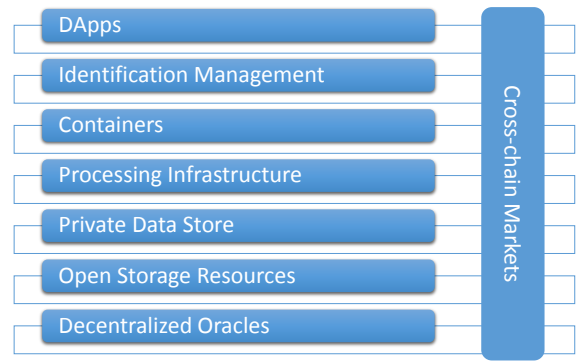

Fig. 4. Layers for the distributed cloud stack

### A. Software Layers for Distributed Clouds

The promise of 3G DLT that introduce sufficient scalability and cross-chain transfers is that it will enable a new distributed infrastructure to take form that can autonomously communicate cross-chain. Contemplating innovation diffusion [31] we can consider that cryptocurrencies have already escalated from the innovator segment to early adopters (in some countries such as Korea perhaps into early majority). Although the early application landscape is currently forming [31], the 2G smart contracts are trapped in the innovation segment as the technology required for their widespread dissemination is limited. Based on project release estimates, such as Ethereums PoS (2018-19) and sharding (2019-20) we can expect that 3G DLT enabled solutions are available in the near future (1-2 years). In addition to enabling a widespread dissemination of smart contract use, 3G DLT will facilitate a software layer stack that can be utilized for building distributed clouds in the medium term (2-5 years), see e.g. Golem project roadmap[8].

Considering the previous discussion of the generational shifts that have occurred and are occurring in distributed ledger technology, we here present a typical stack that may form distributed clouds in the coming decade, see figure 4. This stack should not be seen as definite or obligatory, but rather, as for centralized cloud computing, it depends on implementation and needs. DApps are today composed of front-ends and smart contract(s), this will extend to include more complex layers in the future. Below we present the layers and provide examples of projects implementing specific layers.

- A common way of provisioning **DApp front-ends** is to use a distributed publicly available storage resource such as IPFS or Swarm.
- **Identification management** can be handled by a global identity network such as the proposed Sovrin, that will offer a self-sovereign identity to users. For single-chain bound solutions identification can often be handled through normal chain identification.

[8]http://bit.ly/2v0RR7h

- Data that should be publicly available can be stored in an **open storage resource** e.g. IPFS or Swarm. These platforms also allow for the storage of encrypted data blobs. The protocols used for storage are p2p-based, but with added functionality, e.g. permanence (unique content addressing).
- **Private data stores** on top of a open storage resource can offer individuals a way to store personal data or it may represent a limited access data store such as a private cloud bucket or database (cf. with Filecoin).
- Although the EVM may offer some basic processing capabilities and logic implementations, a **processing infrastructure** will be needed to run more processing intensive tasks, similar to centralized cloud computing IaaS and PaaS. The Golem project proposes a compute model where underutilized resources can be offered by and to anyone, thereby increasing cost efficiencies. Fine-grained FaaS solutions discussed in Section II would be options here as well and provide the opportunity to unify the centralized cloud computing world with the distributed ledger world.
- Heavy and complex processing loads such as analytics workflows can be migrated through a **container** like encapsulation to the processing infrastructure. Streamr has proposed an analytics engine that is able to handle streaming data and possibly a DApp for designing an analytics pipeline.
- **Decentralized oracles** will offer an ability to solve decision problems that need consensus, such as agreeing on an outcome. Reaching consensus on real-world events that then is evaluated by smart contracts can be done through oracles. For the Ethereum platform Augur offers a oracle contract solution for defining and predicting event outcomes based on peer information.
- 3G DLT solutions will offer an ability to transfer monetary value through **cross-chain markets**. This also enables the different layers to coexist and build on each others technology, as payments can be made fully automated and transparent.
- Although resource discovery can be distributed, currently it seems unlikely that e.g. a distributed DNS will offer any technical benefits (it may offer access benefits). Additionally reverse proxy solutions used for static resource requests (cf. CloudFlare) enable ultra low latencies that solutions presented here can hardly compete with. Potentially, edge computing resources may in the future offer, through geographic proximity resources, competing solutions.

Monetization of decentralized platforms are usually performed through the use of tokens. The pricing model for processing-based tasks is likely to follow that for FaaS services, i.e. execution-based or the IaaS/PaaS model, i.e. time activated. For storage, the pricing model should capture the data size, time stored, level of replication, data transfers, and potential data broker fees. By autonomously handling pricing on some ledger-based market we can assume that it remains efficient and transparent for users. Eventually, it is likely that market dynamics will regulate pricing through supply and demand of resources. A secondary pricing model that may become important in the future is Quality of Service (QoS), provided a proof can be constructed to measure QoS in a transparent way. Here we consider for example the proximity to users that often is the basis for edge and fog computing, but also other measures such as trust and access to specialized hardware.

*B. Unifying cloud provisioning approaches*

In Figure 5 three different architectural representations are presented to illustrate ways to construct software for distributed clouds. The leftmost representation is the **DApp design**. The center design makes use of one or several layers (as presented in previous section) of the distributed cloud stack to create the platform and a likely DApp for interaction. The rightmost representation offers access to a layer through a DAPI. The DAPI implementation is yet to be taken into widespread use, first application is likely a payment DAPI for micro transactions. The centermost design represents the utilization of the corresponding PaaS model for centralized cloud computing. The different layers forming the Paas offering will develop over coming years, but an example can be an analytics container engine that targets a specific problem. The advantage of the distributed cloud and PaaS model is that the platform monetization principle is built into the transactional fabric of the ledger concept, with the result that the monetization becomes more transparent compared to platforms built on centralized technology and processing silos [32].

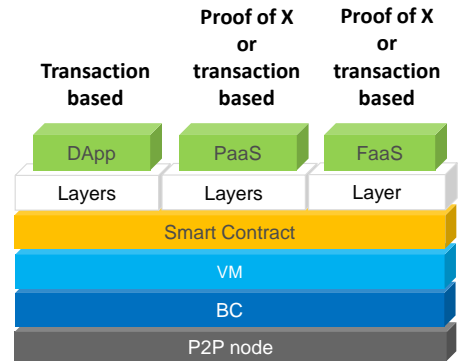

Fig. 5. An Unified Architecture

The technological complexity of utilizing and building distributed PaaS and distributed FaaS should not exceed the centralized counterparts. Rather the transactional nature of the ledger may make it easier to innovate concerning multi-sided markets such as application stores. The inherent facilitation of transactions and/or proofs may make the matching of buyers and sellers and the automated clearance of transactions between the parties truly bilateral [33].

## V. CONCLUSION

This paper reviews the conceptually centralized service provisioning model and distributed ledger technologies in parallel. Their conceptual differences arising from centralised control to distributed and built-in trust, constitute a paradigm shift in software engineering. Still, the similarities in service provisioning are many and in addition, the security weakness of one can help improve the other.

Our initial contribution can be found in the service provisioning definition of cloud computing (cf. sec. II) and the review of the maturity in distributed ledger technology (cf. sec. III). From this we then envision a future path were infrastructure control become obsolete and access to the compute stack is seen as a pure utility (cf. sec. IV). We elaborate on a definition for distributed clouds and their service provision offerings in relation to incumbent conceptually centralised solutions.

Our main contribution is in the identification of a potential paradigm shift in the architectural design of the distributed cloud and the mechanisms enabling a shift towards distributed clouds, while maintaining the present service provisioning architecture. This, a novel attempt at bridging the gap between two seemingly competing and different cloud models, indicate that the commonalities in terminology for cloud users may be more abundant than what first meets the eye.

Distributed cloud solutions may in the beginning be considered less secure. Centralized cloud computing has gone through a similar path, and the question of processing sensitive data in the centralised cloud is still debated. Although DLT has security built into its core design (which is not normally the case for centralized cloud services) the EU General Data Protection Regulation adds further considerations for the purpose of processing personal data on rented infrastructure [34] that needs further elaboration. Particularly in regards to forensic aspects of securing audit trail data. A great number of security aspects also require further study. A key problem will be data leakage from and possibly data manipulation in p2p processing nodes.

Still, economical effectiveness of distributed clouds, self-auditing platform mechanics, and a guaranteed availability may eventually turn companies into utilizing distributed cloud solutions.

ACKNOWLEDGMENT

This research is partly funded by German Federal Ministry of Education and Research (Cloud TRANSIT, 13FH021PX4).